# Radio y Redes Cognitivas

*White paper – AEI eMOV*

El objetivo de este *whitepaper* es presentar un conjunto de retos y líneas de investigación prioritarias en relación con el desarrollo de radios y redes cognitivas en los próximos años.

Coordinadores: Carles Antón Haro y Lorenza Giupponi
Correo electrónico: {carles.anton,lorenza.giupponi}@cttc.es

23/03/2010

# ÍNDICE

# Executive Summary

Traditionally, two different policies to access the radio spectrum have coexisted: licensed regulation, whereby the rights to use specific spectral bands are granted in exclusivity to an individual operator; or unlicensed regulation, according to which certain spectral bands are declared open for free use by any operator or individual following specific rules. While these paradigms have allowed the wireless communications sector to blossom in the past, in recent years they have evidenced shortcomings and given signs of exhaustion. For instance, it is quite usual to encounter fully overloaded mobile communication systems coexisting with unused contiguous spectral bands. This clearly advocates for a more flexible and dynamic allocation of the spectrum resources which can only be achieved with the advent of the so-called cognitive radios and networks.

This whitepaper provides an accurate description of priority research activities and open challenges related to the different functionalities of cognitive radios and networks. First, we outline the main open problems related to the theoretical characterization of cognitive radios, spectrum sensing techniques as well as the optimization of physical layer functionalities in these networks. Second, we provide a description of the main research challenges that arise from a system point of view: MAC protocol optimization, traffic modelling, RRM strategies, routing paradigms or security issues. Next, we point out other problems related to the practical hardware implementation of cognitive radios, giving especial emphasis to sensing capabilities, reconfigurability and cognitive control and management. Finally, we succinctly report on a number of current activities related to the standardization of cognitive radio systems.

Despite that research in cognitive radios and networks has spurred lots of interest, many open issues at the scientific, technological and regulatory levels, still remain this preventing from a widespread adoption of cognitive radio and network technologies. This calls for an increased and more coordinated research work which includes (i) to spend further efforts to understand and characterize the performance limits of cognitive radios from an information theoretical perspective, along with the design of advanced physical layer and spectrum sensing techniques; (ii) to investigate novel methods for an effective management of the aggregated interference generated by the secondary users (based e.g. on artificial intelligence techniques, cooperative transmission, geo-located databases or the use of cognitive pilot channels); (iii) to develop specific security mechanisms for cognitive networks; (iv) to encourage a closer interaction between the signal processing, network protocol, radiofrequency and microelectronic design communities in relation with the design of reconfigurable radio platforms; (v) to closely monitor developments in and actively contribute to a number of European and international regulatory bodies (ETSI, ITU-R, IEEE); and, finally, (vi) to increase the support from publicly-run and private research funding programs in order to underpin the aforementioned research and standardization efforts.

# LISTA DE ACRÓNIMOS

| ACRÓNIMO | SIGNIFICADO |
|---|---|
| BBDD | Bases de Datos |
| CDMA | Code Division Multiple-Access |
| CEPT | European Conference of Postal and Telecommunication administrations |
| CN | Cognitive Networks |
| CogNeA | Cognitive Network Alliance |
| CPC | Cognitive Pilot Channel |
| CR | Cognitive Radio |
| DoS | Denegación de Servicio (*Denial-of-Service*) |
| DSP | Digital Signal Processing |
| DVB-T | Digital Video Broadcasting - Terrestrial |
| ECMA | European Computer Manufacturers Association |
| ETSI | European Telecommunication Standards Institute |
| FCC | Federal Communications Commission |
| FPGA | Field Programmable Gate Array |
| GPS | Global Positioning System |
| IEEE | Institute of Electrical and Electronics Engineers |
| IEEE SCC41 | IEEE Standards Coordinating Committee 41 |
| ISO | International Organization for Standardization |
| ITU | International Telecommunication Union |
| LTE | Long Term Evolution |
| MAC | Medium Access Control |
| NC4 | Network-Coded Cognitive Control Channel |
| OFA | Object Function Attacks |
| OFCOM | Office of Communications |
| OFDMA | Orthogonal Frequency Division Multiple Access |
| OSI | Open System Interconnection |
| PUE | Primary User Emulation |
| QoS | Quality of Service (Calidad de Servicio) |
| RF | Radiofrecuencia |
| SCC | Standard Coordinating Committee |
| SDR | Radio definida por Software (Software-Defined Radio) |
| SEAMCAT | Spectrum Engineering Advanced Monte Carlo Analysis Tool |
| TC RRS | Technical Committee for Reconfigurable Radio Systems |
| TDT | Televisión Digital Terrestre |
| UHF | Ultra-High Frequency |
| WGSE | Working Group Spectrum Engineering |
| WG | Working Group (Grupo de trabajo) |
| WRC | World Radiocommunication Conference |

# 1 Introducción[1]

En lo referente a la regulación del espectro radioeléctrico, dos han sido las políticas que de manera tradicional se han venido aplicando: la asignación a operadores de derechos exclusivos de uso sobre algunas bandas frecuenciales (bandas licenciadas), y la declaración de otras como bandas de uso libre por parte de operadores y/o usuarios particulares (bandas no licenciadas). Si bien es cierto que una aplicación equilibrada de dichas políticas ha permitido un desarrollo vigoroso del sector de las telecomunicaciones, algunos hechos recientes como el elevado coste que han alcanzado las licencias en los procesos de subasta, la percepción de que el espectro radioeléctrico es un bien escaso, o la baja eficiencia espectral de los esquemas de acceso al medio en bandas no licenciadas (debida en gran medida a la ausencia de coordinación entre usuarios y a los elevados niveles de interferencia), sugieren la necesidad de considerar otras políticas complementarias.

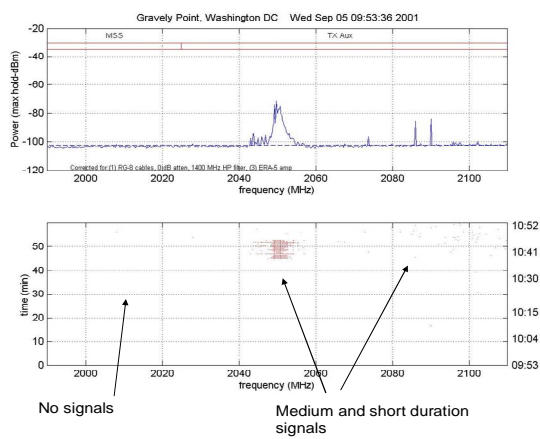

**Figura 1: Ocupación espectral medida en Washington DC por la empresa Shared Spectrum Company.**

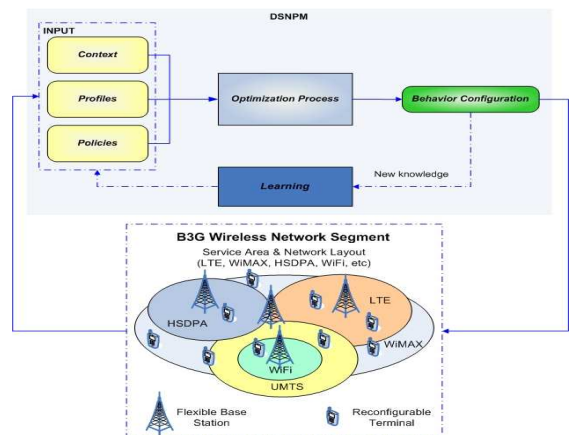

**Figura 2: Mecanismos de gestión en redes cognitivas (gentileza del proyecto E3, financiado por la Comisión Europea).**

El principal problema radica en que las políticas actuales de asignación de bandas frecuenciales son demasiado estrictas y no permiten un uso óptimo del espectro radioeléctrico disponible. Por ejemplo, a pesar de que el espectro se percibe como un bien escaso, el uso de buena parte de las bandas licenciadas es muy bajo o nulo. A consecuencia de ello, no es infrecuente encontrar bandas frecuenciales terriblemente congestionadas (como, por ejemplo, las de telefonía móvil celular) y bandas contiguas completamente infrautilizadas. La Figura 1 pone de manifiesto este problema. En ella se muestra la ocupación del espectro frecuencial en Washington DC un día laborable cualquiera a las 10:00 de la mañana (medidas llevadas a cabo por la empresa Shared Spectrum Company, www.sharedspectrum.com). En el gráfico inferior, se detalla la evolución temporal de la ocupación dicha banda frecuencial y de él se desprende que (1) una parte muy significativa del espectro no se usa; y (2) que la ocupación, cuando se produce, es a ráfagas, es decir, con señales intermitentes de baja y media duración. En resumen, el uso del espectro asignado de manera estática puede ser muy ineficiente lo que, en última instancia, podría acarrear serias dificultades a la hora de desplegar nuevos sistemas de comunicaciones inalámbricas.

---

[1] Este documento está basado en parte en el *whitepaper* titulado "Cognitive Radio Platforms" de la Plataforma Tecnologica Europea eMobility. En su redacción se ha hecho especial hincapié en enfatizar e incluir un mayor grado de detalle sobre los retos tecnológicos y prioridades a nivel nacional expresados por empresas, universidades y centros de investigación.

Todo ello sugiere que deberían explorarse otras políticas más flexibles y dinámicas de asignación de recursos radioeléctricos como las que se derivan del uso de las denominadas radios cognitivas (CR, del inglés *Cognitive Radios*). Los dispositivos cognitivos tienen la capacidad de medir la ocupación espectral de determinadas bandas frecuenciales, determinar cuáles no estan siendo utilizadas y reconfigurar de manera apropiada sus parámetros de transmisión (forma de onda, potencia, etc.) para aprovechar dichos huecos espectrales. Tales conceptos pueden extrapolarse a las denominadas redes cognitivas (CN, del inglés *Cognitive Networks*) entendidas como redes que son capaces de aprender del uso que se hace de ellas en determinadas circunstancias para, en última instancia, adaptar su comportamiento en aras de un uso más eficiente de las mismas. De manera general, la información de entrada de la que disponen las CN para llevar a cabo tal adaptación puede clasificarse en *información contextual*, *perfiles de uso*, y un conjunto predefinido de *políticas de reconfiguración* (ver Fig. 2). La información contextual, como por ejemplo la ubicación de los terminales, el volumen de tráfico generado, nivel de movilidad o de interferencia, etc., debe ser continuamente monitorizada por la red. Por otra parte, los *perfiles de uso* proporcionan información sobre las diferentes configuraciones posibles de los elementos de la red, a nivel, por ejemplo, de tecnologías radio (3G, Wi-Fi, etc.), códigos o portadoras disponibles. Por último, las *políticas de reconfiguración* hacen referencia a las reglas que deben seguirse a la hora de seleccionar los anteriormente mencionados *perfiles de uso* en función de la *información contextual* disponible. A menudo, las *políticas de reconfiguración* persiguen maximizar la calidad de servicio (QoS, del inglés *Quality of Service*) ofrecida al usuario (en términos de velocidad de transmisión, cobertura o fiabilidad de la red) y/o minimizar los costes que ello le genera al operador. Tales políticas acostumbran a diseñarse en base a técnicas de inteligencia artificial o aprendizaje máquina.

La reconfiguración de la redes cognitivas puede llevarse a cabo en todos los niveles (capas) de la pila OSI de protocolos. A nivel físico y de control de acceso al medio (MAC, del inglés *Medium Access Control*), dicha reconfiguración puede suponer el ajuste de la tecnología de acceso radio (RAT, del inglés *Radio Access Technology*) a emplear, la banda frecuencial, la potencia transmitida o el esquema de modulación. A nivel de red, los cambios pueden afectar a los protocolos de enrutamiento utilizados, los mecanismos de control de la congestión o las topologías de red (mallada, en estrella, etc.). Por último, a nivel de aplicación y de servicio es posible, por ejemplo, modificar los niveles de QoS asignados a dichas aplicaciones.

## 2 Líneas de Investigación

El objetivo de esta sección no es otro que el de discutir un conjunto de retos y lineas de investigación prioritarias en relación con el desarrollo de radios y redes cognitivas en los próximos años. La clasificación se ha llevado a cabo en función de la capa de la pila de protocolos OSI a la que hacen referencia. Así, la sección 2.1 aborda un conjunto de problemas abiertos relacionados con la caracterización teórica y la optimización de la capa física de dichas redes. A continuación, la sección 2.2 se centra en otros aspectos a nivel de sistema como el modelado de tráfico, o las necesidades de mejora de los protocolos MAC, la gestión de recursos radio, los esquemas de enrutamiento o los mecanismos de seguridad dentro del contexto de redes cognitivas. De manera complementaria, la sección 2.3 describe otros retos relacionados con la implementación de radios cognitivas. Para finalizar, la sección 2.4 da una visión de conjunto de los diferentes procesos de estandarización relacionados con radios y redes cognitivas que se están llevando a cabo en la actualidad.

## 2.1 Diseño de capa física para radios cognitivas y caracterización a nivel de teoría de la información

La teoría de la información proporciona valiosas herramientas de cara a la caracterización de los límites teóricos y el comportamiento asintótico de los sistemas de comunicaciones. Así, las redes cognitivas pueden modelarse como un caso particular del denominado canal con interferencias (*interference channel*) en donde el objetivo del usuario secundario o cognitivo es transmitir la mayor cantidad de información posible a su destinatario sin perturbar en exceso al usuario primario o licenciado. En dicho proceso, es crucial que el usuario secundario haga el mejor uso posible de toda aquella información que, sin estar destinada a él, pueda tener conocimiento por haber haber sido transmitida por el usuario primario; o de cualquier conocimiento previo sobre la estructura de la señal interferente. En este contexto, es imprescindible el estudio, caracterización y diseño de nuevas arquitecturas de transmisión y recepción que hagan un uso efectivo de dicho conocimiento *a priori* de la interferencia. La extensión de dichas arquitecturas y técnicas a escenarios con un número arbitrario de usuarios primarios y secundarios es un reto que también debe ser abordado.

El sensado del espectro, definido como la búsqueda de huecos espectrales (*spectrum holes*) mediante la exploración del espectro radioeléctrico en las proximidades de un receptor, es una funcionalidad clave para el correcto funcionamiento de las radios cognitivas. Una vez detectados, los huecos espectrales deben ser utilizados por los usuarios secundarios para realizar sus emisiones sin interferir a los usuarios licenciados. Las tareas fundamentales en sensado del espectro son las siguientes: (1) detección de huecos espectrales; (2) estimación de la resolución espectral de cada hueco; (3) estimación de las direcciones espaciales de las interferencias; y (4) clasificación de las señales. A nivel de estado de arte, se ha propuesto ya un gran número de técnicas de procesado de señal y aprendizaje máquina basadas en distintas propiedades de las señales de los usuarios primarios: energía, correlación temporal o espacial, cicloestacionariedad, pilotos disponibles, etc. A pesar de ello, el problema del sensado del espectro está todavía lejos de estar definitivamente resuelto pues se enfrenta a desafíos enormes entre los que cabe destacar la naturaleza aleatoria del canal inalámbrico, o las incertidumbres relativas a la disponibilidad de los huecos espectrales pues poco se conoce sobre la estadística de su aparición y desaparición. Por todo ello, es necesario desarrollar nuevas técnicas de sensado del espectro que sean rápidas, adaptativas y fiables, en particular para señales de bajo nivel de potencia. A tal efecto, el uso de técnicas de sensado cooperativas, donde varios terminales secundarios geográficamente dispersos colaboran en la detección de la señal del usuario primario, constituye una técnica prometedora. Algunas líneas de investigación relevantes en este campo incluyen la coordinación distribuida y eficiente de las secciones de espectro a sensar por cada nodo, el diseño de las reglas de detección espectral y fusión de dichas medidas, o las estrategias cooperativas de distribución y fusión de la información de sensado. Por otra parte, el hecho de que la monitorización espectral deba realizarse a menudo sobre grandes bandas frecuenciales (monitorización de banda ancha) plantea un nuevo conjunto de retos. Recientemente el concepto de muestreo compresivo (*compressed sensing*) ha supuesto una auténtica revolución en lo que se refiere a la adquisición y muestreo de datos analógicos en un esfuerzo hacia resolver la problemática de recuperar un proceso continuo comprimible con un nivel suficiente de similitud si únicamente se realiza un número muy reducido de medidas o muestras del mismo. La aplicación de este tipo de técnicas al sensado espectral propio de los sistemas de radio cognitiva supone una línea de investigación de gran auge e interés investigador. Asimismo, la implementación práctica de dispositivos que realicen el muestreo compresivo en banda ancha requiere de una fuerte inversión de esfuerzo investigador a corto y medio plazo.

Como se comentaba en el párrafo anterior, el conocimiento en tiempo real de la ocupación espectral en un determinado entorno geográfico debe ser utilizado por las radios cognitivas para minimizar el impacto de las interferencias generadas por ellos sobre los usuarios licenciados. Ello es posible mediante el uso de las denominadas técnicas de adaptación de la forma de onda basadas, por ejemplo, en sistemas con múltiples antenas (mediante las cuales se puede minimizar la potencia radiada en la dirección del usuario primario), o en modulaciones multiportadora, siendo de especial interés, a tenor del elevado grado de selectividad frecuencial, aquellas basadas en técnicas de bancos de filtros. El diseño conjunto de la codificación/compresión de fuente y codificación de canal también puede ser aplicado a la adaptación de la modulación a las características de la señal de información a transmitir. El objetivo no es sólo la mejora de las prestaciones extremo a extremo (e.g. tasa de error de bit, complejidad computacional en transmisión), sino también la minimización de la interferencia que usuarios no licenciados imprimen en las señales de usuarios primarios en escenarios *spectrum underlay*. En escenarios de optimización intercapa (como el anteriormente mencionado de codificación conjunta de fuente y canal) sumamente complejos de resolver mediante métodos y herramientas numéricas convencionales, la aplicación de esquemas de optimización heurística permite ajustar la bondad de las soluciones subóptimas obtenidas a la carga computacional asumible. Una de los mayores retos en este campo es el desarrollo de alternativas a los denominados algoritmos genéticos.

Por último, la implementación práctica de radios cognitivas requiere el intercambio en tiempo real de una cantidad considerable de información acerca del estado del canal. Es importante estudiar qué información debe intercambiarse y cómo comprimirla para no producir un overhead excesivo que podría saturar la red de radio cognitivas. Debe estudiarse el diseño y la optimización de los canales de señalización (como el denominado *cognitive pilot channel,* CPC, propuesto por el proyecto europeo E$^3$) que envían la información del canal y que habitualmente tienen una tasa muy limitada. Asimismo, otra cuestión importante es que los canales de señalización proporcionan información imprecisa acerca del canal radio debido en parte a que está desfasada por su naturaleza variante en el tiempo y a errores en la estimación y compresión de dicha información. Es importante que las funcionalidades de una radio cognitiva tengan en cuenta la imprecisión de esta información y exhiban un comportamiento robusto frente a ellas.

Como resumen de lo dicho en párrafos anteriores, en lo referente a radio cognitiva es absolutamente necesario priorizar las siguientes líneas de investigación:

- Caracterización de sistemas de radio cognitiva desde un punto de vista de la teoría de la información. Dicha caracterización permitirá establecer los límites teóricos en lo que se refiere a las prestaciones que dichos sistemas cognitivos pueden ofrecer (tasa de bit, eficiencia espectral, etc.) así como valiosas reglas de diseño para los mismos.

- Desarrollo de nuevas técnicas de estimación espectral rápida, fiable y de banda ancha. A tal efecto, es necesario avanzar en el estudio de técnicas avanzadas de sensado colaborativo y muestreo compresivo.

- Diseño de técnicas avanzadas de modulación adaptativa de forma de onda, que permitan adaptar las características de la señal transmitida por los usuarios secundarios a los huecos espectrales existentes. Ello incluye, entre otros, el diseño de esquemas de modulación multiportadora, filtrado espacial y codificación conjunta de datos y canal.

- Diseño y optimización de los canales de señalización que envían la información contextual y del canal.

## 2.2 Redes cognitivas

Esta sección se centra en las líneas de investigación más prometedoras en el ámbito de las redes cognitivas.

La funcionalidad clave en las radios cognitivas es el control de la interferencia en el receptor primario, que se convierte en un problema sin fácil solución, en escenarios donde dicho receptor es pasivo, como es el caso de sistemas de televisión. El objetivo último es garantizar el correcto funcionamiento de los usuarios licenciados y al mismo tiempo maximizar el rendimiento de los usuarios secundarios. Una manera eficaz de abordar este problema es utilizar técnicas de transmisión cooperativa entre usuarios secundarios lo que permitiría maximizar su tasa de transmisión empleando niveles de potencia muy bajos y sin causar interferencias. De manera similar, se estudia la aplicabilidad de conceptos de cooperación entre nodos en entornos cognitivos, donde los usuarios secundarios compensan a los usuarios primarios del espectro por la interferencia causada a sus receptores, a través de cooperación. La cooperación se implementa a través del reenvío al receptor primario de mensajes primarios interceptados por los transmisores secundarios debido a la naturaleza *broadcast* del entorno inalámbrico. El receptor primario podrá así mejorar la robustez de su proceso de detección, gracias a la redundancia de los mensajes recibidos, y a su diversidad, debido a los diversos canales radio a que éstos están expuestos.

Otra alternativa para garantizar la coexistencia entre los sistemas primarios y secundarios a nivel de interferencia es utilizar una red de sensores auxiliares que recojan la información de los usuarios licenciados y controlen las transmisiones de los usuarios secundarios. Esta información se podría, por ejemplo, proporcionar a una base de datos a la cual los secundarios accederían antes de realizar transmisiones en banda licenciada. Esta opción, representa una alternativa muy interesante a la detección por medio de sensado de espectro, y conlleva la implementación y el mantenimiento de bases de datos que indique en qué frecuencias está permitida la operación de terminales cognitivos, en qué localización y cuándo y cómo pueden usarse. Dicha información es transmitida a los terminales cognitivos con el objeto de que éstos puedan acceder al espectro de forma oportunista y sin interferencias. Este método resulta de especial interés en escenarios con geolocalización, es decir, donde la posición de los terminales licenciados y de los terminales cognitivos se conoce o se puede aproximar con poco margen de error. Las bases de datos geolocalizadas conllevan una serie de problemáticas (método de geolocalización, precisión máxima e impacto en términos de interferencia causada al sistema licenciado, etc.), que deben ser abordadas y cuya resolución propiciará un mayor desarrollo de las redes cognitivas en este ámbito de aplicación. De particular interés resulta el estudio del efecto de la movilidad en estos escenarios, la geolocalización en entornos tanto interiores como exteriores, el diseño de las entidades, funcionalidades y arquitectura relacionadas con el mantenimiento y funcionamiento de la base de datos (arquitecturas centralizadas o distribuidas), los protocolos de señalización y acceso para la difusión de información referente a espectro disponible, la robustez a posibles ataques de denegación de servicio (DoS, del inglés *Denial of Service*), la periodicidad con que debe actualizarse la base de datos, o la aplicabilidad en entornos TDT (Televisión Digital Terrestre) y redifusión. Este tipo de solución más conservadora respecto a la que se basa en el sensado del espectro, es la que los órganos reguladores ingles y americano, OFCOM y FCC (*Federal Communication Commission*), consideran actualmente como más viable para el uso secundario del espectro.

Otra línea de investigación más agresiva que se lleva adelante en diferentes proyectos europeos, como ARAGORN (*Adaptive reconfigurable access and generic interfaces for optimisation in radio Networks* - http://www.ict-aragorn.eu/), $E^3$ (*End-to-End Efficiency* - https://www.ict-e3.eu/) y BeFEMTO (*Broadband evolved Femtocells* – http://www-ict-befemto.eu/) es el control adaptativo de la interferencia mediante algoritmos distribuidos de inteligencia artificial. Las redes neuronales, por ejemplo, representan una tecnología prometedora para la implementación de las capacidades de aprendizaje de las radios cognitivas aunque algunos aspectos como el tiempo de aprendizaje, los requisitos de memoria o la complejidad computacional deben ser investigados de manera más detallada. Por otra parte, en escenarios donde múltiples nodos deben tomar decisiones de manera distribuida es necesario aplicar algoritmos de aprendizaje multiagente. El escenario cognitivo se mapea fácilmente en un sistema multiagente, por las siguientes razones: 1) cada radio cognitiva tiene información parcial sobre su entorno; 2) no existe necesariamente una entidad centralizada que gestione los recursos de la red secundaria; 3) los estímulos que las radios cognitivas tienen en consideración en la toma de decisión, provienen de fuentes de información distribuidas espacialmente en el escenario; 4) los procesos de decisión de las diferentes radios son asíncronos en el tiempo; 5) la decisiones independientes de las diferentes radios dependen de las decisiones de las otras radios y también les afectan. Algunos problemas que quedan por resolver en este area de conocimiento son la optimización de decisiones tomadas de modo descentralizado o el análisis de los comportamientos oscilatorios mediante herramientas basadas en teoría de juegos. Una posible solución que mitigaría estas oscilaciones y mejoraría el proceso de aprendizaje, consiste en explotar la cooperación entre nodos. Por ejemplo, un nodo más inteligente o que ha tenido más oportunidades para aprender estrategias eficientes de toma de decisiones, podría elevarse a "entrenador" o "docente", para otros nodos con capacidades de aprendizaje más limitadas. Este paradigma, conocido como *docitive radio*, (del latín, *docere* = enseñar y *cognoscere* = conocer) emula el paradigma social de estudiante-docente y puede proporcionar beneficios significativos en términos de tiempo de aprendizaje y eficiencia del proceso de toma de decisiones.

Otro problema que no ha sido completamente resuelto es el acceso eficiente al medio en redes constituidas por dispositivos inalámbricos equipados con transceptores capaces de sintonizar un sólo canal a la vez. Una solución prometedora pasa por el uso de canales de control cognitivos basado en esquemas de codificación de red (NC4, del inglés *Network-Coded Cognitive Control Channel*) que permite a las diferentes radios cognitivas saltar entre diferentes canales siguiendo secuencias pseudoaleatorias e intercambiar información de control cada vez que se encuentran en el mismo canal. Mediante el uso de técnicas de codificación de red, la información de control puede ser distribuida de manera eficiente entre los diferentes nodos que configuran dicha red. Esta información se puede utilizar de forma que dichos nodos puedan coordinarse eficientemente para acceder a porciones del espectro radio infrautilizadas, sin la necesidad de la implementación de un canal común de control.

La seguridad es otro aspecto clave para el desarrollo de las redes cognitivas, tanto en entornos civiles como militares (dentro del estándar IEEE 802.22 se contempla, por ejemplo, un grupo específico de trabajo dedicado a esta temática). Entre los principales tipos de ataque para los cuales es necesario investigar posibles contramedidas cabe destacar los siguientes:

- Ataques OFA (del inglés, *Object Function Attacks*), destinados a afectar a los algoritmos de aprendizaje de los dispositivos de radio cognitiva.

- Ataques "*cross-layer*", entre la capa física/enlace y la capa de transporte, destinados a afectar al protocolo de transporte saturando la red.

- Ataques PUE (del inglés, *Primary User Emulation*), en los que se imita la transmisión producida por un elemento primario para evitar la transmisión en la misma banda por parte de un elemento secundario.

- Ataques de sensado. En general, en el caso de redes cognitivas basadas en el paradigma de sensado cooperativo, un posible atacante puede introducir información falsa con el objetivo de afectar a la operatividad general de la red.

Una primera aproximación a estos problemas pasa por el establecimiento de sistemas de reputación de terminales. Debe observarse, sin embargo, que estos sistemas deben seguir unos requisitos exigentes de eficacia y eficiencia, debido a la gran cantidad de interacciones necesarias entre dispositivos y a las características de movilidad en el espacio y en la utilización del espectro por su parte.

Respecto a la arquitectura del sistema cognitivo, la investigación tiende a concentrarse en el ámbito de las arquitecturas descentralizadas, donde los nodos actúan de manera autónoma y autoorganizada. Cuando la densidad de nodos transmisores y receptores es elevada, es práctica habitual que un número reducido de ellos actúe como concentradores, debiendo el resto asociarse a dichos nodos. En el escenario más simple, dicha asignación debe ser llevada a cabo de tal modo que los nodos concentradores sean capaces de soportar la tasa de datos agregada de sus nodos asociados. El problema se complica si se introducen restricciones adicionales tales como el nivel de interferencia entre nodos, la prioridad entre nodos licenciados y no licenciados, o el ajuste de potencia y modulación en sistemas multiportadora. Las denominadas técnicas heurísticas permiten establecer compromisos entre la complejidad computacional y la bondad de las posibles soluciones.

Finalmente, la implantación de un mercado secundario del espectro conlleva necesariamente el desarrollo de sistemas de información en tiempo real sobre el espectro utilizado (frecuencias utilizadas y disponibles, modulaciones empleadas, ámbito espacial de aplicación, etc.), el modelado y abstracción de las estaciones base desplegadas así como de los correspondientes terminales de usuario (frecuencias de trabajo, modulaciones, tecnologías compatibles, tasas de transferencia, retardos y latencias en función de la banda y tecnología usada, geolocalización); o la caracterización de servicios a proporcionar (tasas de transferencia necesarias, tolerancia a retardos e interrupciones de sesión, necesidades de autenticación, etc).

Resumiendo, es crucial invertir esfuerzos en las siguientes líneas de investigación:

- Métodos de gestión de la interferencia agregada generada por múltiples radios cognitivas, ya que casi diez años después de la introducción del concepto de radio cognitiva, la interferencia generada por el sistema secundario, aún representa la mayor preocupación de los reguladores y de los propietarios del espectro.

- Aplicación de técnicas de inteligencia artificial distribuida: En particular, es necesario evaluar la aplicabilidad de dichos algoritmos, en términos de coste computacional, requisitos de memoria, y velocidad de aprendizaje en entornos inalámbricos altamente cambiantes.

- Mecanismos de seguridad específicos para redes cognitivas: La seguridad es aún un problema abierto tanto desde el punto de vista de la investigación, como en estandarización. El IEEE 802.22 ha dedicado algunos esfuerzos en esta línea, pero todo indica que aún queda muchos problemas por resolver.

- Diseño de canales de control común: Definir la manera en la que las radios cognitivas se coordinan entre ellas e intercambian información es tambien una prioridad para la realización de las redes cognitivas.

## 2.3 Plataformas y testbeds para radios y redes cognitivas

Una radio cognitiva está constituida por cuatro unidades fundamentales: (1) una unidad de sensado que proporciona información respecto al entorno, en términos, por ejemplo, de ocupación del espectro; (2) un mecanismo cognitivo que estructura la información sensada y puede tener conocimiento de criterios complemetarios como, por ejemplo, políticas de gestión del espectro, máscaras de espectro, etc.; (3) una unidad de decisión que determina cuando y cómo la radio debe reconfigurarse para adaptarse a los cambios del entorno;

y (4) una unidad de radio flexible que modifica los parámetros de transmisión y recepción en base a las decisiones tomadas por la unidad de decisión.

Para beneficiarse plenamente del concepto de radio cognitiva, es absolutamente necesario disponer de plataformas radio flexibles. Sin embargo, las plataformas radio actualmente disponibles en el mercado no resultan adecuadas, por ejemplo, para realizar de manera eficiente labores de escaneado frecuencial y detección de bandas ocupadas y libres (fundamentalmente, en términos de velocidad, coste y gasto energético). Además, los nuevos algoritmos de procesado de la señal, basados en detectores de energía o en propiedades cicloestacionarias, son capaces de encontrar un compromiso entre la fiabilidad del sensado y el nivel de conocimiento que el sistema secundario tiene que tener del sistema primario o la complejidad computacional. Por este motivo, es necesario desarrollar plataformas radio que sean capaces de soportar la implementación de dichos algoritmos. Asimismo, es fundamental dar cabida también a nuevas arquitecturas que permitan realizar el sensado de múltiples canales en paralelo sin perder flexibilidad en lo referente a su capacidad de modulación de la frecuencia central.

Respecto a la reconfigurabilidad en transmisión y recepción, tanto el sistema de banda base digital, como el cabezal de radiofrecuencia analógico deben ser capaces de adaptarse a las distintas formas de onda. Los requisitos son particularmente estrictos: es necesario poder pasar rápidamente de un canal al otro, obtener altas prestaciones a baja potencia y tener capacidad de cancelación de interferencias. Una opción para la implementación de la plataformas radio reconfigurables es la denominada radio definida por software (SDR), que asume que todos los módulos de la radio se implementan vía software. Idealmente, la señal recibida debería muestrearse directamente en radiofrecuencia, es decir, realizando la conversión analógico-digital justo después de la antena, con el reto tecnológico que ello supone. La investigación en plataformas SDR se centra en tres aspectos principales: (1) el diseño de plataformas hardware abiertas basadas en elementos de software configurable como FPGA (del inglés, *Field Programmable Gate Array*), microprocesadores, y DSP (del inglés, *Digital Signal Processing*); (2) la implementación de la forma de onda en tiempo real, para que ésta se pueda mapear en la plataforma, para las diferentes tecnologías de comunicación; y (3) el desarrollo de mecanismos de reconfigurabilidad de la radio, que a menudo se basan en virtualización del hardware a través de un *middleware* o software intermedio. De manera alternativa y más realista a corto y medio plazo, otras plataformas radio reconfigurables se diseñan en base a componentes parametrizables, lo que permite adaptar su evolución a la de las capacidades de los componentes electrónicos.

Las estrategias de control y gestión cognitivas se ocupan de aspectos relacionados con la operación en tiempo real de los dispositivos inalámbricos cognitivos. Esto incluye el diseño de funcionalidades avanzadas de gestión para que los dispositivos inalámbricos determinen de manera automática su configuración y comportamiento en base a perfiles, políticas, experiencias y conocimiento. En otras palabras, los dispositivos cognitivos deben estar provistos de las siguientes capacidades: (1) gestión de las preferencias y comportamientos del usuario, capacidades del dispositivos y políticas de red, como por ejemplo modelos de tráfico, patrones de movilidad, nivel de calidad del servicio preferido por el usuarios para ciertas aplicaciones, etc; (2) adquisición de la información de contexto, como el chequeo periódico de la disponibilidad de las diferentes tecnologías de acceso radio, o la monitorización de estadísticas relevantes; (3) decisión de la red y tecnología de acceso más apropiada y selección de la acción de reconfiguración más adecuada.

Además del diseño de las plataformas radio, es necesaria la creación de entornos de pruebas y Living Labs para sistemas de radio cognitiva. Estos entornos deben atender a criterios específicos de la tecnología, de manera que se garantice la viabilidad y aplicabilidad de las pruebas a realizar. En concreto deberán estudiarse los entornos de aplicación en busca de escenarios que (1) abarquen entornos en los que la densidad de uso del espectro, tanto libre como licenciado, sea relevante para el estudio de viabilidad de la solución propuesta; (2) presenten una utilización significativa y habitual de sistemas en bandas licenciadas con terminales tanto activos (telefonía móvil, etc.) como pasivos (televisión, radio, etc.); (3) permitan el estudio en condiciones tanto estáticas como en movilidad; y (4) comprendan a toda la cadena de valor en la aplicabilidad de la tecnología, desde el usuario final hasta el proveedor último de servicios sobre ella.

Finalmente, el <u>desarrollo de herramientas de simulación que permitan evaluar el impacto de la interferencia generada por los sistemas secundarios</u> es de gran importancia. Un ejemplo es la herramienta SEAMCAT[2], basada en el método de simulación de Monte Carlo, que ha sido desarrollada dentro del marco del grupo de trabajo WGSE (*Working Group Spectrum Engineering*) del CEPT (*European Conference of Postal and Telecommunication administrations*). En este sentido, los desarrolladores de la herramienta se están planteando también la inclusión de elementos de detección de espectro para dar soporte a terminales cognitivos que acceden al espectro compartido de forma no interferente. Este hecho y la amplia caracterización de escenarios (CDMA, OFDMA, etc.) y modelos de propagación, así como el carácter abierto del código fuente, convierten a esta herramienta en firme candidata para el desarrollo de una plataforma de radio cognitiva con garantías. Algunos tareas de investigación a realizar en base a esta plataforma incluyen la evaluación de estrategias de acceso cognitivo basadas en bases de datos geolocalizadas (cuantificar la cantidad de información necesaria a distribuir a los terminales cognitivos como bandas de frecuencia libres, potencia máxima, etc.) o la investigación de alternativas para el uso compartido de espectro en los planos frecuencial, temporal, espacial, etc. que conduzcan a un aprovechamiento espectral máximo en entornos de aplicabilidad como ofrece la TDT, LTE-Advanced, y otros. Para ello, sin embargo, es necesario afrontar algunos retos pendientes como el desarrollo interfaces que permitan el postprocesado de los resultados de simulación con otras herramientas (p.e. MATLAB®).

Como conclusión, los sistemas cognitivos constituyen una tecnología extremadamente prometedora, pero su eventual desarrollo se ve comprometido por problemas de implementación, tanto desde el punto de vista hardware, como software, así como por la disponibilidad de plataformas radio flexibles y simuladores. Por todo ello, es <u>crucial investigar en las siguientes áreas</u>:

- <u>Diseño conjunto de plataformas y sistemas reconfigurables</u>: Esto es esencial para garantizar que los avances en sistemas inalámbricos relacionados con sistemas cognitivos, puedan probarse en sistemas y entornos reales.

- <u>Desarrollo de esquemas de reconfigurabilidad novedosos</u>: A modo de ejemplo, se destaca el concepto de las radios definidas por software, que podrían ser capaces de generar soluciones a largo plazo, aunque su implementación hoy en día dista mucho de ser una realidad.

---

[2] http://www.ero.dk/seamcat

- <u>Estrategias para el control y la gestión de las plataformas y sistemas cognitivo</u>: De hecho, los beneficios de la radio cognitiva dependen de cómo la radio se comporta, así que control y gestión adecuados son esenciales.

- <u>Desarrollo y mejora de herramientas de simulación</u> como, por ejemplo, la herramienta SEAMCAT. El objetivo en última instancia es poder evaluar el impacto de la interferencia generada por los sistemas secundarios sobre los primarios.

## 2.4 Actividades de Estandarización

La radio cognitiva abre nuevas vías de acceso al espectro radioeléctrico por lo que, desde el punto de vista de los órganos reguladores, constituye a la vez una oportunidad de aumentar la eficiencia en el uso de dicho espectro, así como un herramienta valiosa para el despliegue de nuevos servicios. Al mismo tiempo, es necesario garantizar los derechos de los propietarios actuales del espectro y, por este motivo, la introducción en el mercado de la radio cognitiva debe venir precedida del correspondiente proceso estandarizador. El objetivo último es identificar nuevas políticas de gestión del espectro, conjuntos de indicadores de prestaciones que permitan garantizar la protección de los usuarios primarios frente a los secundarios, así como posibles limitaciones técnicas que puedan dificultar la coexistencia de diferentes sistemas en la misma banda.

Respecto a los <u>principales actores</u> en el ámbito de la estandarización cabe destacar:

- La ITU (*International Telecommunication Union*) que, en 2007 y con vistas a la *World Radiocommunication Conference* del año 2012 (WRC-12), desarrolló un apartado especifico al respecto del desarrollo de los sistemas cognitivos. Se ocupa de dicho apartado el grupo de trabajo ITU-R WP1B dedicado a "metodologías para la gestión del espectro y estrategias de mercado".

- En el ámbito de ETSI (*European Telecommunication Standards Institute*), el comité técnico para sistemas radio reconfigurables (del inglés TC RRS) tiene el encargo de redactar informes de viabilidad definiendo qué estándares relacionados con radio cognitiva deben desarrollarse. En particular, este comité se encarga de definir los requisitos de los principales propietarios de espectro, y está organizado en 4 grupos de trabajo (WG): WG1 Aspectos de sistema, WG2 Arquitectura del equipamiento radio; WG3 Arquitectura funcional y canal de control cognitivo; WG4 Seguridad pública.

- El IEEE *Standards Coordinating Committee* 41 (IEEE SCC41, antes IEEEP1900) se centra en las redes cognitivas para acceso dinámico del espectro. El objetivo es desarrollar estándares para redes cognitivas de futura generación. Algunos ejemplos representativos del trabajo realizado en el marco del SCC41 son la definición del estándar para sensado espectral y las estructuras de datos para el acceso dinámico del espectro (IEEE P1900.6); o los estándares para optimizar el uso de los recursos radios en redes heterogéneas caracterizadas por múltiples RATs. En este contexto, los dispositivos son terminales con capacidades cognitivas que acceden a múltiples RATs tomando decisiones de manera descentralizada y operando de manera flexible en diferentes bandas de frecuencia (IEEE P1900.4).

- Diferentes asociaciones internacionales sin ánimo de lucro trabajan también en el ámbito de la radio cognitiva. La más representativa es el SDR Forum cuyos socios son actores de ámbito comercial, militar, gubernamental, proveedores de servicio, fabricantes, agencias de regulación, universidades y centros de investigación. De manera complementaria, la CogNeA (del inglés, *Cognitive Network Alliance*) agrupa a diferentes actores del ámbito de la electrónica de consumo interesados en atraer a diferentes fabricantes de equipos hacia la producción de dispositivos personales portátiles de banda ancha y bajo consumo capaces de operar en las banda televisiva de UHF (del inglés, *Ultra High Frequency*)

- Finalmente, la ECMA (*European Computer Manufacturers Association*) es otra asociación de industrias relacionadas con las tecnologías de la información que a menudo contribuye a los órganos de estandarización como ETSI o el ISO (*International Organization for Standardization*).

La situación actual a nivel regulatorio es como sigue. En Europa se están valorando de manera exhaustiva las posibilidades de la radio cognitiva y sus mecanismos de uso de la mano de grupos de trabajo como SE43 en CEPT o el grupo de trabajo RSS de ETSI. En algunos casos, se trata de una respuesta a mandatos de la Comisión Europea en este sentido, siendo uno de los objetivos el poder anticipar las decisiones que previsiblemente se tomarán en la próxima conferencia mundial de la ITU-R de Enero de 2012 (WRC-12). En contraposición, en Estados Unidos se ha venido desarrollando un trabajo más intenso que ha desembocado en la decisión de la FCC (de Noviembre de 2008) por la que se autoriza el uso de dispositivos cognitivos en los canales libres de la banda de televisión. Igualmente, y dado que la orden de la FCC requería que los dispositivos cognitivos accediesen a una base de datos informando de su posición para así recabar la información de canales libres, está en marcha el proceso de selección de aquellas entidades que gestionarían dicha base de datos. Cabe esperar que, del mismo modo, en Europa se priorize el uso de bases de datos geolocalizadas frente a otros esquemas de monitorización espectral. En cualquier caso, los estándares ya desarrollados o en proceso de estarlo, tales como el IEEE 802.22 o los de la asociación ECMA, deberían ser modificados para su adaptación al entorno europeo en todo aquello que tiene que ver con los niveles de potencia y detección espectral.

Respecto a futuras asignaciones frecuenciales, la porción de espectro entre 790 y 862 MHz (canales 61 a 69), también conocida como dividendo digital, se va a asignar a servicios móviles a partir de 2015 en toda Europa. Esto conlleva la liberación de dichos canales de servicios de difusión de televisión lo que, en algunos casos como el Español, supondrá una reordenación importante de canales asignados a la distribución de TDT. Por otro lado, los servicios de seguridad y emergencias se encuentran con una porción escasa de espectro, con escasa armonización y diferentes estándares. La radio cognitiva puede suponer una oportunidad siempre que se logre garantizar la presencia de huecos espectrales en cualquier ubicación geográfica.

A la vista de lo anterior, sería deseable priorizar en España las siguientes líneas de actuación en lo relacionado con actividades de estandarización:

- Aprovechamiento de los canales 21 a 68 con el objeto de facilitar la existencia de claros espectrales (*spectrum holes*) que puedan ser reutilizados por otros servicios. En este sentido, la existencia de al menos un canal libre a lo largo de toda la geografía española, no necesariamente en la misma ubicación, podría servir para su uso por parte de servicios de seguridad y emergencia, complementando las bandas actuales destinadas a tal efecto.
- <u>Puesta en marcha de la Agencia Estatal de Radiocomunicaciones</u>, cuya creación está contemplada en la Ley 32/2003 General de Telecomunicaciones, para centralizar y reforzar todas las actividades relacionadas con la gestión del espectro, impulsando además la participación española en las actividades de estandarización internacionales.
- <u>Determinación del papel de los operadores primarios en el uso cognitivo de su espectro</u>: oportunidades de negocio, estrategias de señalización (faros), ajuste dinámico de su potencia en función del uso de su espectro, etc.

Por otro lado, <u>a nivel internacional es muy importante invertir esfuerzos en</u>:

- <u>Armonizar la terminología y definición de modelos de referencia</u>: Hay una profunda necesidad de dar coherencia a la terminología y definir modelos de referencia apropiados para la radio cognitiva y la radio definida por software. Es inmediato constatar como ambos conceptos se utilizan a menudo de manera inapropiada, o ambigua y, lo que es peor, el hecho de que poseen significados distintos para distintas comunidades científicas. A modo de ejemplo, la radio cognitiva, para algunos grupos se refiere al acceso *dinámico* del espectro, para otros al acceso *oportunista* a dicho espectro, y para otros a radios inteligentes, conscientes del entorno. Resulta evidente, por tanto, la necesidad de armonizar dichos conceptos.
- <u>Contribuciones relevantes a organismos de estandarización</u>: Es necesario, por ejemplo, redoblar los esfuerzos para entender las posibles aplicaciones de los claros espectrales. CEPT ha empezado a trabajar en este sentido, con la creación del grupo de ingeniería del espectro 43 (SE43 – *Spectrum Engineering Group* 43), para la banda UHF. Cabe espera que las conclusiones a que se llegue sean determinantes para el futuro de la radio cognitiva.

## 3 RECOMENDACIONES

La puesta en marcha de sistemas cognitivos con un modelo de negocio que los respalde no es todavía una realidad, ni siquiera en Estados Unidos donde se ha autorizado el uso de la banda de televisión por dispositivos con capacidades cognitivas aunque sometidos a fuertes restricciones. La falta del marco legal apropiado junto con un estado tecnológico incipiente frenan, de momento, la puesta en el mercado de soluciones. Sin embargo, el gran potencial que presenta esta tecnología para un uso más eficiente del espectro nos lleva a proponer las siguientes <u>recomendaciones</u>:

R1) Profundizar en el conocimiento y la caracterización de los límites teóricos alcanzables por los dispositivos cognitivos desde un punto de vista de la teoría de la información, así como avanzar en el diseño de técnicas de sensado (p.e. para señales DVB-T) y de capa física imprescindibles para un correcto despliegue de la radio cognitiva.

R2) Investigar en nuevos métodos para la gestión de la interferncia agregada generada por múltiples radios cognitivas mediante técnicas distribuidas de inteligencia artificial, transmisión cooperativa, el empleo de bases de datos geolocalizadas y el diseño eficiente de canales comunes de control, entre otros. Donde proceda, desarrollar nuevas herramientas de simulación que permitan evaluar el impacto de la interferencia generada por los usuarios secundarios.

R3) Desarrollar mecanismos de seguridad específicos para redes cognitivas.

R4) Incrementar el grado de colaboración entre investigadores con conocimientos en las áreas de procesado de la señal, diseño de protocolos de capa de enlace y red, y diseño de dispositivos de radiofrecuencia, en aras de resolver los diversos problemas tecnológicos asociados con el desarrollo de plataformas radio reconfigurables así como mecanismos realistas de sensado del espectro.

R5) Mejorar el seguimiento y contribución española a los diferentes organismos de estandarización europeos (CEPT) e internacionales (ITU-R), tratando de impulsar marcos legales armonizados que sirvan de impulso a la tecnología.

R6) Reforzar el apoyo a nivel nacional en lo referente a políticas regulatorias favorables a la radio cognitiva (asignación de canales, órganos reguladores, etc.)

R7) Aumentar el volumen de financiación, a nivel nacional e internacional, de los programas de investigación públicos y privados destinados a apoyar el desarrollo de la radio y redes cognitivas, tanto a nivel tecnológico como investigador.

# Listado de redactores

| NOMBRE | ENTIDAD | CORREO |
|---|---|---|
| **Carles Anton Haro** | CTTC | carles.anton@cttc.es |
| **Luis Castedo Ribas** | Universidad de A Coruña | luis@udc.es |
| **Javier del Ser Lorente** | Fundación ROBOTIKER TECNALIA-TELECOM | jdelser@robotiker.es |
| **Armin Dekorsy** | Qualcomm | adekorsy@qualcomm.com |
| **Miguel Egido Cortés** | Grupo GOWEX | megido@gowex.com |
| **Xavier Gelabert** | iTEAM | xavier.gelabert@ieee.org |
| **Lorenza Giupponi** | CTTC | lorenza.giupponi@cttc.es |
| **Xavier Mestre** | CTTC | xavier.mestre@cttc.es |
| **Jose Monserrat** | iTEAM | jomondel@iteam.upv.es |
| **Carlos Mosquera** | Gradiant | mosquera@gradiant.org |
| **Miquel Soriano** | UPC | soriano@entel.upc.es |
| **Liesbet van der Perre** | IMEC | Liesbet.VanderPerre@imec.be |
| **Jon Arambarri** | Fundación Universitaria Iberoamericana | jon.arambarri@funiber.org |
| **Juan Antonio Romo** | Univ. del País Vasco | jtproarj@bi.ehu.es |